\documentclass[aps,prl,twocolumn,superscriptaddress]{revtex4}

\usepackage{graphicx}
\usepackage{amsmath}
\usepackage{color}

\begin{document}

\title{Observation and origin of the $\Delta$-manifold in Si:P $\delta$-layers}
\author{Ann Julie Holt}
\affiliation{Department of Physics and Astronomy, Interdisciplinary Nanoscience Center (iNANO), University of Aarhus, 8000 Aarhus C, Denmark}
\author{Sanjoy K. Mahatha}
\affiliation{Department of Physics and Astronomy, Interdisciplinary Nanoscience Center (iNANO), University of Aarhus, 8000 Aarhus C, Denmark}
\author{Raluca-Maria Stan}
\affiliation{Department of Physics and Astronomy, Interdisciplinary Nanoscience Center (iNANO), University of Aarhus, 8000 Aarhus C, Denmark}
\author{Frode S. Strand}
\affiliation{Department of Physics, Center for Quantum Spintronics, Norwegian University of Science and Technology (NTNU), Trondheim, Norway}
\author{Thomas Nyborg}
\affiliation{Department of Physics, Center for Quantum Spintronics, Norwegian University of Science and Technology (NTNU), Trondheim, Norway}
\author{Davide Curcio}
\affiliation{Department of Physics and Astronomy, Interdisciplinary Nanoscience Center (iNANO), University of Aarhus, 8000 Aarhus C, Denmark}
\author{Alex Schenk}
\affiliation{Department of Physics, Center for Quantum Spintronics, Norwegian University of Science and Technology (NTNU), Trondheim, Norway}
\author{Simon P. Cooil}
\affiliation{Department of Physics, Center for Quantum Spintronics, Norwegian University of Science and Technology (NTNU), Trondheim, Norway}
\affiliation{Department of Mathematics and Physics, Aberystwyth University, Aberystwyth, United Kingdom}
\author{Marco Bianchi}
\affiliation{Department of Physics and Astronomy, Interdisciplinary Nanoscience Center (iNANO), University of Aarhus, 8000 Aarhus C, Denmark}
\author{Justin W. Wells}
\affiliation{Department of Physics, Center for Quantum Spintronics, Norwegian University of Science and Technology (NTNU), Trondheim, Norway}
\author{Philip Hofmann}
\affiliation{Department of Physics and Astronomy, Interdisciplinary Nanoscience Center (iNANO), University of Aarhus, 8000 Aarhus C, Denmark}
\author{Jill A. Miwa}
\affiliation{Department of Physics and Astronomy, Interdisciplinary Nanoscience Center (iNANO), University of Aarhus, 8000 Aarhus C, Denmark}
\email{miwa@phys.au.dk}

\date{\today}
\begin{abstract}
By creating a sharp and dense dopant profile of phosphorus atoms buried within a silicon host, a two-dimensional electron gas is formed within the dopant region. Quantum confinement effects induced by reducing the thickness of the dopant layer, from $4.0$\,nm to the single-layer limit, are explored using angle-resolved photoemission spectroscopy. The location of theoretically predicted, but experimentally hitherto unobserved, quantum well states known as the $\Delta$-manifold is revealed. 
Moreover, the number of carriers hosted within the $\Delta$-manifold is shown to be strongly affected by the confinement potential, opening the possibility to select carrier characteristics by tuning the dopant-layer thickness.
\end{abstract}
\maketitle

The process of $\delta$-doping is to create a high-density doping profile within a narrow, well-defined region of a semiconductor. By creating a $\delta$-layer of phosphorus in a silicon host, a strong potential is induced in the dopant layer region, giving rise to a highly conductive two-dimensional electron gas (2DEG) \cite{Polley2012, Miwa2013}. $\delta$-layers of this kind are the structural element behind a significant number of intriguing developments towards a scalable silicon-based solid state quantum computer, such as the first single-atom transistor \cite{Fuechsle2012}, the narrowest conducting nanowire \cite{Weber2012}, an atomically precise tunnelling junction \cite{House2014}, and the fabrication of spin qubits \cite{Koch2019, Wang2016}. 
The arrangement of a single or few phosphorus atoms act as hosts for spin qubits, whereas larger dopant regions form the basis of source, drain and gate electrodes. 

For these reasons, considerable effort has been dedicated to developing a complete understanding of Si:P $\delta$-layers, in particular the factors influencing their electronic structure. Due to the challenges of measuring buried electron states, theoretical models based on tight-binding (TB) and density functional theory (DFT) have dominated the field \cite{Carter2009, Carter2011, Lee2011, Drumm2012, Carter2013, Smith2014}. These theoretical calculations predict metallic quantum states forming both at the centre and close to the corners of the surface Brillouin zone (SBZ) of Si, known as the $\Gamma$- and $\Delta$-states respectively. The $\Gamma$-states are the most occupied, and several of these states are predicted to exist below the Fermi level ($E_F$), depending on the degree of phosphorus doping \cite{Mazzola2019}. Accurate theoretical predictions of the energy splitting between these states, \textit{i.e.} the so-called valley splitting, are challenging and have resulted in values ranging from 6--270\,meV \cite{Drumm2013}, making experimental verification a necessity. Previously, angle-resolved photoemission spectroscopy (ARPES) measurements confirmed the formation of the $\Gamma$ quantum states at the Brillouin zone (BZ) centre \cite{Miwa2013}, as well as a description of their orbital character \cite{Mazzola2014}, phonon and impurity interactions \cite{Mazzola2014a}, and the aforementioned valley splitting of the $\Gamma$-states \cite{Miwa2014}. While a comprehensive description of the $\Gamma$-states is thus underway, the theoretically predicted metallic states closer to the SBZ corners, the so-called $\Delta$-manifold have not been experimentally observed, leaving the theoretical calculations in qualitative disagreement with the observations by ARPES. Since the $\Delta$-states are found near the SBZ corners, and four-fold degenerate, they are expected to have a high impact on the density of states at $E_F$. The presence of additional quantum states near $E_F$ would also have crucial implications for Si:P $\delta$-layer based qubit systems, since a variety of excited states will be possible by the many configurations of valence electrons within the $\Delta$-manifold \cite{Fuechsle2010}.

In this Letter, we confirm the existence of the  theoretically predicted $\Delta$-manifold using high-resolution ARPES.  The states are located at the corners of the Si SBZ, \textit{i.e.} at $\textbf{k}=[\pm 0.68, \pm 0.68]$\,\AA\textsuperscript{-1}, determined with an uncertainty of $\pm 0.02$\,\AA\textsuperscript{-1}.  Their symmetry and location is in agreement with theoretical models \cite{Carter2009, Carter2011, Lee2011, Drumm2012, Carter2013, Smith2014}, settling the incongruity between theory and experimental measurements. In addition, the effect of quantum confinement is investigated as the P dopant layer is reduced from a 4\,nm thickness down to a single-atom-thick layer limit. (Note that we use single-layer (SL) to refer to the single-atom-thick layer limit.)
We find that the $\Gamma$- and $\Delta$-states have a qualitatively different response to a modified confinement potential, resulting in a redistribution of carriers between the quantum states. This behaviour opens the possibility for selecting carrier characteristics by tuning the dopant-layer thickness; a capability which could be capitalized on to enhance the performance of atomic-scale devices constructed from Si:P $\delta$-layers.

\begin{figure}
\includegraphics{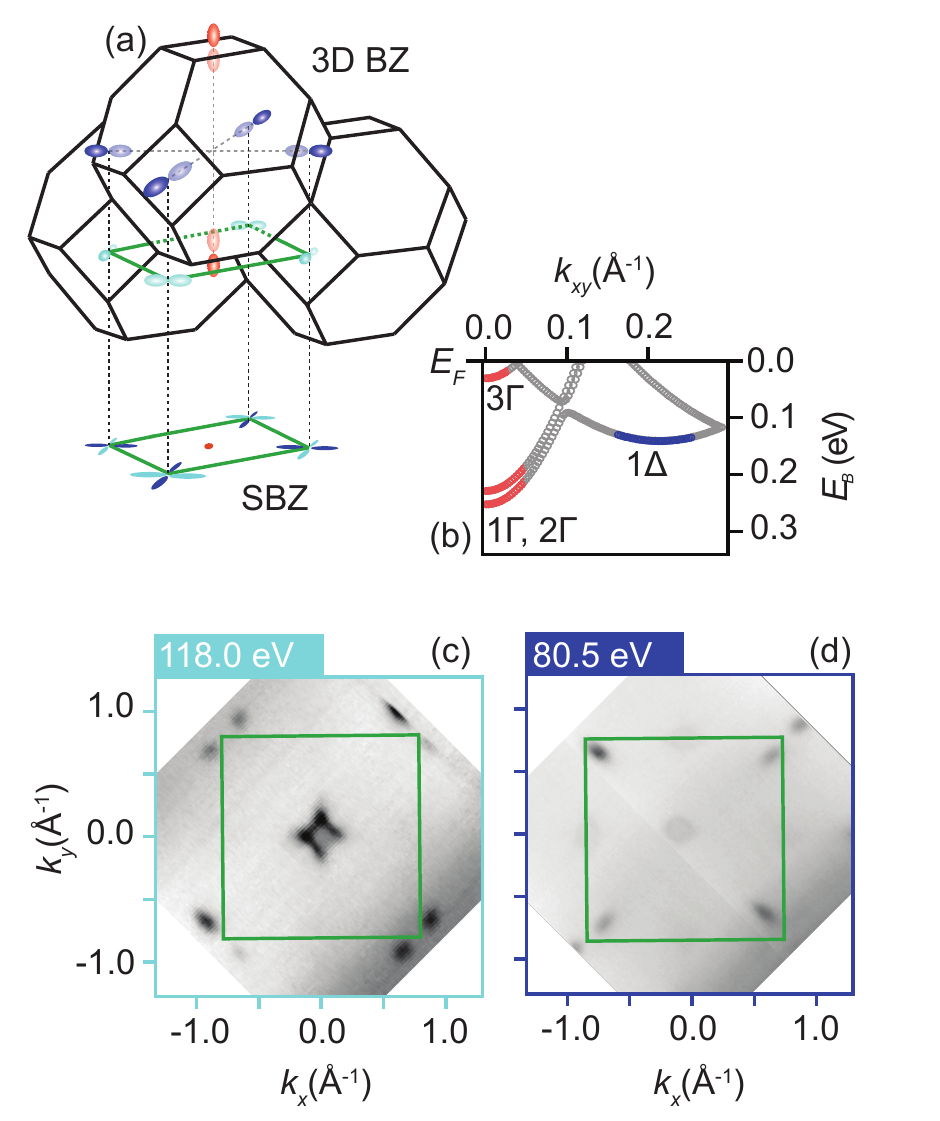}\\
\caption{(a) Projection diagram for a highly confined $\delta$-layer. Out-of-plane CBM valleys (red) are projected onto the SBZ centre, whereas in-plane CBM valleys (blue) are projected close to the SBZ corners. (b) TB band structure calculation predicting the formation of quantum well states (adapted from \citet{Mazzola2019}). The most occupied bands ($\Gamma$-states) originate from confinement of the out-of-plane CBMs, whereas the in-plane CBMs result in shallow $\Delta$-states. (c) Constant energy surface at $E_F$ acquired with 118.0\,eV and (d) 80.5\,eV photons. By combining the intensity of the two spectra all the quantum states depicted in (a) are accounted for in the ARPES data.}
  \label{fig:DFT}
\end{figure}

Three adjacent three-dimensional (3D) BZs of bulk Si are illustrated in Figure \ref{fig:DFT}(a). In the 3D BZ at the forefront, the six conduction band minima (CBMs) are shown. Projecting the 3D BZ and the electronic structure onto the (001) surface results in a square SBZ (green) and conduction band minimum (CBM) derived electron pockets near the centre and the corners of the SBZ.
The reduced BZ, shown in Figure \ref{fig:DFT}(b), is calculated using an empirical $sp^3d^5s^*$ TB model coupled with the Poisson equation for a 2$\times$2 supercell. The modelled band structure is adapted from \citet{Mazzola2019} which follow the calculations described in \citet{Lee2011}.
Several confined bands are predicted to appear below $E_F$, conveniently available for photoemission spectroscopy. The most occupied quantum bands, referred to as the $\Gamma$-states, arise from the out-of-plane CBMs (red), projected onto the zone centre. The $\Gamma$-states are split because of the formation of bonding and antibonding states as they end up at the same $k_\parallel$. Projection of the in-plane CBMs (blue) result in quantum states appearing close to the SBZ corners, known as the $\Delta$-manifold. Each 1$\Delta$-state is located at a distinct value of $k_\parallel$ and therefore not split in energy.


\begin{figure*}
\includegraphics{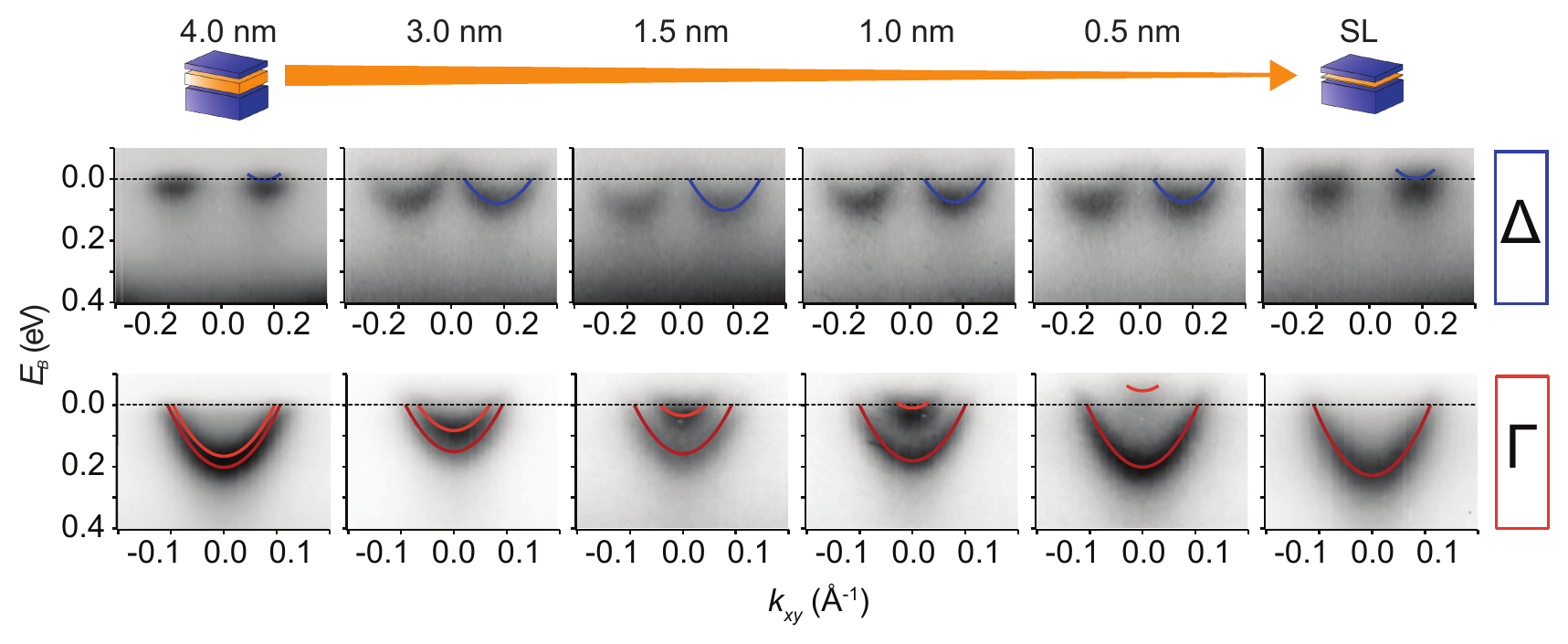}\\
\caption{Energy dispersion showing the development of the 1$\Delta$-states as the dopant-layer thickness is reduced (upper row). Corresponding development for the $\Gamma$-states (lower row). The measurements are acquired with 44.0\,eV and 37.0\,eV photons for the $\Delta$- and $\Gamma$-states, respectively. As the confinement increases, the fitted bands of the $\Gamma$-states separate in energy, whereas the 1$\Delta$-band initially moves closer to 1$\Gamma$ in terms of energy, before reversing this behaviour upon reaching a dopant-layer thickness of 1.5\,nm.}
  \label{fig:AllSamples}
\end{figure*}

Fabrication of Si:P $\delta$-layer samples was made following a known procedure, which yields a $\frac{1}{4}$- monolayer (ML) of P dopants on a clean Si(001) surface buried beneath epitaxially grown Si \cite{Miwa2013, Miwa2014, Mazzola2014}. Thicker dopant layers were grown by co-deposition of PH$_3$ ($\textit{i.e.}$ the P dopant source) and Si to produce samples with a dopant concentration on par with the SL case; see \citet{Mazzola2019} for details. In this work, the dopant layers were buried beneath $1.5\pm0.5$\,nm of epitaxial Si. ARPES measurements were acquired at the SGM3 beamline at the ASTRID2 synchrotron (Aarhus, Denmark). During data acquisition the sample was held at room temperature, and the measurements were obtained using a PHOIBOS 150 hemispherical analyzer (SPECS GmbH) with the energy and angular resolutions set to 30\,meV and 0.2\,$^{\circ}$, respectively \cite{Hoffmann2004}.

ARPES measurements showing constant binding energy slices at $E_F$ for a Si:P $\delta$-layer with a 4.0\,nm-thick dopant layer are presented in Figure \ref{fig:DFT}(c) and (d). The measurements were acquired with a photon energy of 118.0\,eV and 80.5\,eV, respectively. In accordance with previous reports, the $\Gamma$-states are observed as intense features appearing at the  centre of the SBZ. The shape of the 1$\Gamma$ Fermi contour is highly anisotropic in comparison to 2$\Gamma$; the bands are flatter in the    $k_{x}$ and $k_{y}$ directions than in the $k_{xy}$ direction \cite{Carter2009, Lee2011, Miwa2014}. In addition to the dominant $\Gamma$-states, other electron states are also clearly present close to the SBZ corners. These states are assigned to the $\Delta$-manifold, and appear in the positions predicated by theory.  Even though all these predicted states are observed, see Figures \ref{fig:DFT}(c) and (d), they are not all observed at the same photon energy. The reason for this deviation is that the quantum confined states still retain some of their $k_{\perp}$ character and their ARPES signal is very weak unless they are resonantly excited \cite{Mazzola2014}.




As the 3D BZ of Si, see Figure \ref{fig:DFT} (a), is a truncated octahedron, the nearest neighbouring 3D BZ is shifted in the $k_{\perp}$ direction. This shift makes the $\Delta$-states appear in a pair-wise fashion at specific values of $k_{\perp}$ in the extended BZ scheme, and their enhanced intensity is therefore obtained pair-wise at separate photon energies. In the ARPES measurement presented in Figure \ref{fig:DFT}(c), there are no  $\Delta$-states within the first SBZ. A pair of $\Delta$-states are, however, visible near each of the SBZ corners but located within the adjacent zones. By reducing the photon energy from 118.0\,eV to 80.5\,eV, see Figure \ref{fig:DFT}(d), the other pair of $\Delta$-states are visible. Together, the two data sets account for all the predicted quantum well states.

The effect of quantum confinement may be investigated by varying the thickness of the dopant layer.  We find that decreasing the dopant-layer thickness from 4.0\,nm down to a SL drastically alters the properties of the quantum states. Measurements obtained from samples with different dopant-layer thicknesses are presented in Figure \ref{fig:AllSamples}. The $\Delta$-states are shown in the upper panels, while the lower panels show the corresponding $\Gamma$-states. The dispersion of the bands is determined by a two-dimensional (2D) fit of the full ARPES spectra, allowing polynomial dispersion up to the third degree for the bare bands. This is a powerful fitting procedure, where one can compare the entire $E$- and $k$-dependent data set to a resolution-broadened model of the spectral function \cite{Nechaev2009, Mazzola2013, Mazzola2014a, Andreatta2019, Hofmann2009}. Further information regarding the 2D fitting process can be found in the Supplementary Material.

Two parabolic bands are used to respresent  the  $\Gamma$-states  (red and orange). The most occupied band hosts both the 1$\Gamma$ and 2$\Gamma$ states, and the less occupied band is assigned to the 3$\Gamma$ state \cite{Mazzola2019}. As the confinement increases the bands separate in energy, and the induced splitting between them becomes more pronounced. Upon reaching the SL limit 3$\Gamma$ is pushed completely above $E_F$, resulting in a lower estimate between the two parabolic bands of 230\,meV for describing the energy splitting. As the $\Gamma$-states are forced to split apart in terms of energy, the increased confinement also affects the binding energy of the $\Delta$-states, see upper panels of Figure \ref{fig:AllSamples}, relative to that of the most occupied $\Gamma$-band. This separation, however, is not monotonically increasing. Experiencing an increase in confinement potential, the $\Delta$-states first shift closer to $1\Gamma$ in terms of energy, before the opposite behaviour is instigated upon reaching a confinement exceeding that produced by a 1.5\,nm-thick dopant layer. 

The divergent dependency on dopant-layer thickness between the separate quantum well bands results in a change of the relative number of carriers within each band, as well as the total number of carriers. The carrier density in the dopant-layer systems is estimated from the occupancy of the bands and presented in Table \ref{tab:CarrierDensity}. At $E_F$, the constant energy contour of 1$\Delta$ is assumed to be elliptical, with an estimated ratio of 1:2.17 between the short and long axis. A four-fold sinusoidal shape is used to describe the Fermi contour of 1$\Gamma$, where the ratio between the short and long radius is estimated to be 1:1.48, whereas 2$\Gamma$ and 3$\Gamma$ are assumed to be isotropic. The shape of these contours are based on the work presented in \citet{Mazzola2019} and the quoted ratios are based on an average determined from constant energy slices, acquired at $E_F$, for the different $\delta$-layer thicknesses. 


\begin{table}
\caption{Total carrier density estimation for the different $\delta$-layer systems together with the relative distribution of carriers within each band. The carrier densities are determined within a maximum absolute uncertainty of $3 \times 10^{13}$ e$^-/$cm$^2$.} 
\label{tab:CarrierDensity} 
\includegraphics{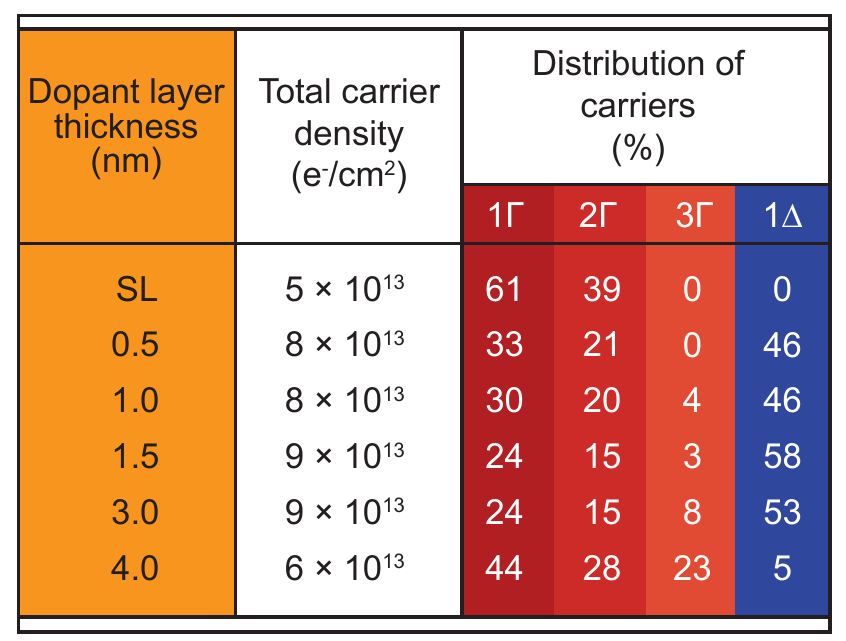}
\end{table}

For the SL doping limit, the electron carrier density is estimated to be $n_{SL}=3.0\times10^{13}$\,e$^-$/cm$^2$, well above the insulator-to-metal transition \cite{Kravchenko2004} and in agreement with earlier transport studies \cite{Goh2006, Goh2008, McKibbin2010}. For this system, $1\Gamma$ and 2$\Gamma$ are solely responsible for hosting electron transport at zero temperature, which may explain the previous elusiveness of the $\Delta$-manifold. The $1\Delta$-states are, however, located close enough to $E_F$ to be thermally populated at room temperature.  


The $\delta$-layer system with a 4.0\,nm-thick dopant layer shows a similar distribution, in the sense that the electron carriers are primarily hosted by the $\Gamma$-bands, located at the BZ centre. Simplified transport calculations only including the $\Gamma$-bands, and may therefore serve as a good model for both these situations. For the intermediate  dopant-layer thicknesses, however, the carrier distribution changes significantly. The carrier contribution from the 1$\Delta$-states can no longer be neglected, since these are found to account for over 46\,\% of the electron carriers in all of the studied intermediate systems, \textit{i.e.} 0.5--3.0\,nm thicknesses. Although the 1$\Delta$-band always appears less occupied than 1$\Gamma$ and 2$\Gamma$, the four-fold degeneracy compensates for the shallow binding energy, giving the $\Delta$-manifold a high impact on the density of states at $E_F$. It is even shown that for a system with a 1.5\,nm-thick dopant layer the 1$\Delta$-states are contributing with 58\,\% of the total carrier density. In order to obtain an accurate description of the electronic properties of Si:P $\delta$-layer-based devices, it is, therefore,  crucial to include the $\Delta$-manifold in any model. 

In summary, the existence of the theoretically predicted $\Delta$-manifold was verified for the first time. The location of these states was shown to be in agreement with DFT and TB calculations, giving strength to the developed models. 
The energy separation between the two parabolic bands used to describe the $\Gamma$-bands was found to increase monotonically with confinement, reaching a separation of at least 230\,meV for the SL limit. The revealed $\Delta$-manifold was shown to accommodate a significant portion of electron carriers. In particular, the $\Delta$-manifold hosts over 46\% of the electron carriers for all systems with dopant-layer thicknesses between the SL and 4.0\,nm. Notably, in the 1.5\,nm-thick dopant-layer sample, 58\% of the carriers are supplied by the $\Delta$-manifold. Such a significant contribution to the carrier statistics demand the inclusion of these states for obtaining an accurate model of any future device based on this platform. In fact, the influence of these states have already been suspected in a study by Fuechsle \textit{et al.} \cite{Fuechsle2010}, where the presence of the $\Delta$-manifold would explain the observed electron states generated in a quantum dot system. The experimental verification and the careful investigation of these states are thus an important step towards obtaining an accurate description of $\delta$-layer-based devices and contribute to the development of a working quantum computer.

\textbf{Acknowledgements} This work was supported by the Danish Council for Independent Research, Natural Sciences under the Sapere Aude program (Grants No. DFF-4002-00029 and DFF-6108-00409), the Aarhus University Research Foundation, and the VILLUM FONDEN via the Centre of Excellence for Dirac Materials (Grant No. 11744), the Research Council of Norway through its Centres of Excellence funding scheme, Project No. 262633, “QuSpin”, and through the Fripro program, Project No. 250985 “FunTopoMat”. Affiliation with the Centre for Integrated Materials Research (iMAT) at Aarhus University is gratefully acknowledged. We also thank C.-Y. Chen and R. Rahman for carrying out the TB calculation presented in this work.

\bibliographystyle{apsrev}

\clearpage

\textbf{Supplementary Material}\\
\vspace{5mm}

\textbf{Details regarding the 2D fitting process}\\

\begin{figure}
\includegraphics[width=0.40\textwidth]{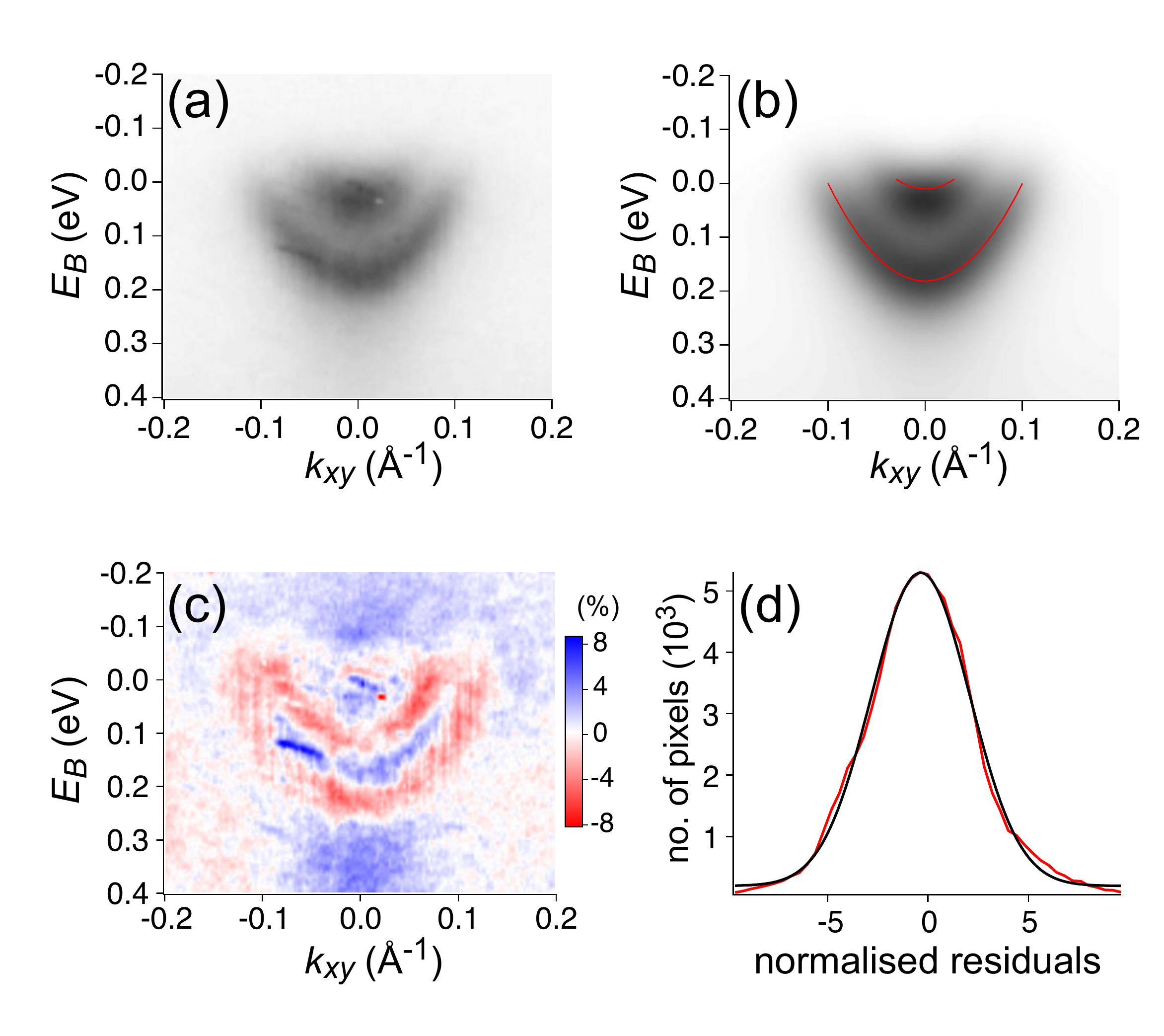}\\
\caption{(a) Example of an ARPES data set. Bare bands are assumed to be well-described by two polynomial functions up to the third degree. (b) Fitted model of data set in (a) with the fitted bare band polynomials overlaid in red. (c) Difference spectrum showing the relative deviation of the fit. (d) Histogram of residuals normalised by pixel error (red curve), fitted to a Gaussian distribution (black curve).}
  \label{fig:2D_fit}
\end{figure}

A conventional approach in determining a band location is to extract line profiles through the measured energy- and momentum-dependent photoemission intensity, \textit{e.g.} energy distribution curves or momentum distribution curves, and fit a model to these profiles. This method may become impractical when the band under investigation is located close, or even above the Fermi level. The distribution of filled electron states will in this case have a crucial impact on the detected band intensity, increasing the complexity of the spectra. Instead of fitting a single line profile one may consider the entire \textit{E}- and \textit{k}-dependent data set (such as in Figure \ref{fig:2D_fit}(a)) and comparing this to a model based on a resolution-broadened spectral function. This procedure is a powerful tool, which may be used to extract information not only about the single-particle dispersion, but also the electronic temperature and many-body interactions \cite{Nechaev2009a, Mazzola2013a, Mazzola2014aa, Andreatta2019a}. 

The ARPES spectral intensity from an ideal measurement is expressed as,
\begin{equation}
\label{eq:Intensity}
I(k,E) \propto  |M_{if}|^2 A(k,E) f_{FD}(E,T).
\end{equation}


\noindent Here, $f_{FD}$ is the Fermi-Dirac distribution, accounting for the fact that only filled electron states are probed, and $|M_{if}|^2$ represents the dipole matrix element. $A(E, k)$ is the hole spectral function, describing the behavior of
an interacting electron system. The spectral function includes the complex self-energy, $\Sigma = \Sigma' + i\Sigma''$, of the interacting electron system by,
\begin{equation}
\label{eq:SpectralFunction}
A(k,E) =  \frac{1}{\pi}\frac{\Sigma''}{(E - E_0(k) - \Sigma')^2 + \Sigma''^2}, 
\end{equation}
\noindent where $E_0(k)$ is the energy of a non-interacting particle, $\Sigma'$ describes the spectral function deviation from the single-particle picture, and $\Sigma''$ gives a measure of the lifetime of the ARPES photohole \cite{Hofmann2009}. A phenomenological model of the photoemission intensity may be expressed as,
\begin{equation*}
\label{eq:IntensityModel}
 \begin{aligned}
I(k,E) = \bigg[ bg(E,k) + \sum_{br} [p_{1,br}(E)+p_{2,br}(k)]\frac{1}{\pi}\\
\frac{p_{5,br}(E)}{(E - p_{3,br}(k)-p_{4,br}(E))^2 + p_{5,br}(E)^2}\bigg] (e^\frac{E}{k_BT}+1)^{-1}.
\end{aligned}
\end{equation*}
\noindent The first term describes an \textit{E}- and \textit{k}-dependent background which are summed with the intensity from any number of branches describing the observed bands. Matrix element variations may give rise to \textit{E}- and \textit{k}-variations in the branch intensity, accounted for by the polynomial functions $p_{1,br}(E)$ and $p_{2,br}(k)$. The single-particle dispersion is described by $p_{3,br}(k)$, and the real and imaginary part of the self-energy is included in $p_{4,br}(E)$ and $p_{5,br}(E)$ respectively, related to each other by the Kramers-Kronig transformation.

It is desirable to implement the above expression with as few fitting parameters as possible that is needed for a satisfactory representation of the observed intensity. In our experiment, the background is found to be sufficiently described by an energy dependent polynomial of second degree. We assume that the real part of the self-energy is zero and that the imaginary part provides a constant broadening of the bare band features. The single-particle dispersion is allowed to be a polynomial of third degree, and the branch intensity is described as independent of both \textit{E} and \textit{k}. In any real measurement the finite resolution in \textit{E} and \textit{k} will affect the acquired spectra, and must be accounted for in the model. This is implemented by a convolution of the ARPES spectral intensity with appropriate resolution functions, G($\Delta$E) and G($\Delta$k), both assumed to be Gaussian.

Figure \ref{fig:2D_fit}(a) presents an example dataset, acquired from the $\Gamma$-states of a Si:P $\delta$-layer sample with a 1.0\,nm-thick dopant layer. A reasonable initial guess for the background parameters and the Fermi level may be obtained by first fitting an area of the spectra where no bands are present. A model of the ARPES intensity is made by providing initial guess coefficients to describe the spectral function, the Fermi-Dirac distribution and the resolution broadening, in addition to the background signal. The fit is performed by varying one or more parameters at a time and comparing the obtained model to the acquired ARPES data. By this procedure, the location of the bare bands may be accurately determined. The resulting model is shown in Figure \ref{fig:2D_fit}(b), with the fitted bare bands overlaid in red.


In order to investigate the accuracy of the fit, the difference spectrum between the model and the acquired ARPES data may be considered. This is presented in Figure \ref{fig:2D_fit}(c), where the colour scale is given in relative deviation from the measured intensity. Modulations in the difference spectrum are present, resulting in blue (high value) and red (low value) regions. A histogram of the fitting residuals, normalised by the error of each pixel, displays a Gaussian distribution with a standard deviation of $\sigma = 2.45$ (a perfect fit within the constraints of the expected noise level implies $\sigma = 1$), shown in Figure  \ref{fig:2D_fit}(d). We have made the approximation that the error contained in each pixel is equal to the square root of the number of counts. The fit deviation is attributed to an oversimplification, in which the reliability of the line-shape is reduced, in favour of a simpler model with less fitting parameters. This does not, however, influence the reliability of the band position or the Fermi level, as these fitting parameters are not significantly correlated to the line-shape parameters.

\bibliographystyle{apsrev}

\end{document}